\documentclass[11pt]{article}
\usepackage{bm}
\usepackage{dsfont}
\usepackage{xcolor}
\usepackage{varwidth}
\usepackage{array}
\usepackage[top=2.5cm, bottom=2.5cm]{geometry}
\usepackage[pdftex,bookmarks=true,bookmarksdepth=4]{hyperref}
\usepackage{pdfpages}
\usepackage{amsmath}
\usepackage{amssymb}
\usepackage{amsfonts}
\usepackage{graphicx}
\usepackage{amsthm}
\usepackage{tikz-cd}
\usepackage{bigints}
\usetikzlibrary{cd}
\usepackage{tikz}
\usepackage{stackengine}
\usepackage{appendix}
\usepackage{mathrsfs}
\usepackage{eso-pic}
\usepackage{titlesec, blindtext, color}
\usepackage[T1]{fontenc}
\usepackage{tocloft}
\usepackage{etoolbox}
\usepackage[nottoc]{tocbibind}
\usepackage{multicol}
\usepackage{wrapfig}
\usetikzlibrary{positioning,backgrounds,decorations.markings}
\usepackage{marginnote}
\usepackage{extarrows}
\usepackage{setspace}
\usepackage{schemata}
\usepackage{mathtools}
\usepackage{changepage}
\usepackage{eso-pic}
\usepackage{bbm}
\definecolor{bluegreen}{rgb}{-.2,.4,0.6}
\definecolor{lightbluegreen}{rgb}{166,225,255}
\definecolor{lightblue}{HTML}{5dbcd2}
\usepackage{tocloft}
\cftsetindents{section}{2em}{3em}
\cftsetindents{subsection}{5em}{3.5em}
\cftsetindents{subsubsection}{8.5em}{4em}

\renewenvironment{thebibliography}[1]{\begin{oldthebibliography}{#1}\setlength{\itemsep}{0.45em}\setlength{\parskip}{0em}}{\end{oldthebibliography}}
\usepackage[some]{background}

\titleformat{\chapter}[hang]{\vspace{0em}\LARGE\scshape\bfseries}{\hspace{-1cm} \thechapter\hsp\hsp\hsp{|}\hsp\hsp\hsp}{0pt}{\LARGE\bfseries}
\titleformat*{\section}{\large\bfseries\scshape}
\titleformat*{\subsection}{\normalsize\bfseries\scshape}
\titleformat*{\subsubsection}{\normalsize\bfseries\scshape}
\titlespacing*{\chapter}{0pt}{25pt}{100pt}
\titlespacing*{\section}{0pt}{15pt}{15pt}
\titlespacing*{\subsection}{0pt}{12.5pt}{12.5pt}
\titlespacing*{\subsubsection}{0pt}{10pt}{10pt}
\theoremstyle{definition}

\usepackage{tocloft}
\cftsetindents{section}{2em}{3em}
\cftsetindents{subsection}{5em}{3.5em}
\cftsetindents{subsubsection}{8.5em}{4em}

\newcommand{\tocseparator}{\addtocontents{toc}{\protect\addvspace{10pt}\hrule\protect\addvspace{5pt}}}
\preto\backmatter{\tocseparator}

\newcommand{\RR}{\mathbb{R}}           % real  numbers
\newcommand{\CC}{\mathbb{C}}
\newcommand{\KK}{\mathbb{K}}

\newcommand{\Lcal}{\mathcal {L}}
\newcommand{\Bcal}{\mathcal {B}}

\newcommand{\Acal}{\mathcal{A}}
\newcommand{\Ecal}{\mathcal{E}}

\newcommand{\Ical}{\mathcal{I}}

\newcommand{\Mcal}{\mathcal{M}}

\newcommand{\Rcal}{\mathcal{R}}

\makeatletter
\DeclareRobustCommand{\tvdots}{\vbox{\baselineskip4\p@\lineskiplimit\z@\kern0\p@\hbox{.}\hbox{.}\hbox{.}}}
\newcommand\dirlim{\mathop{\mathpalette\varlim@{\rightarrowfill@\scriptstyle}}\nmlimits@}
\newcommand\invlim{\mathop{\mathpalette\varlim@{\leftarrowfill@\scriptstyle}}\nmlimits@}
\makeatother
\DeclareFontFamily{U}{mathx}{}
\DeclareFontShape{U}{mathx}{m}{n}{<-> mathx10}{}
\DeclareSymbolFont{mathx}{U}{mathx}{m}{n}
\DeclareMathAccent{\widehat}{0}{mathx}{"70}
\DeclareMathAccent{\widecheck}{0}{mathx}{"71}

\usepackage{extarrows}
\usepackage{setspace}
\usepackage{mathtools}
\usepackage{changepage}

\usepackage{schemata}
\newcommand{\vertiii}[1]{{\left\vert\kern-0.125ex\left\vert\kern-0.125ex\left\vert #1
		\right\vert\kern-0.125ex\right\vert\kern-0.125ex\right\vert}}

\colorlet{grey}{gray!50}
\colorlet{gray}{gray!60}

\definecolor{skin}{HTML}{f9f3de}
\definecolor{wheatskin}{HTML}{c8bea0}
\definecolor{darkskin}{HTML}{83734a}
\definecolor{darkerskin}{HTML}{64521c}

\def\atfield[#1]{\arrow[#1,preaction={draw=white,-,line width=3},nodes={fill=white}]}

\makeatletter
\newcommand\HUGE{\@setfontsize\Huge{28}{0}}
\makeatother
\RequirePackage{fix-cm}
{
\definecolor{bluegreen}{rgb}{-.2,.4,0.6}
\definecolor{lightbluegreen}{rgb}{166,225,255}
\definecolor{lightblue}{HTML}{5dbcd2}
\makeatletter

\def\@seccntformat#1{\csname the#1\endcsname\hspace*{0.5em}$|$\hspace*{0.5em}}
\makeatother
\definecolor{gray55}{gray}{0.55}
\newcommand{\hsp}{\hspace{2.5pt}}
\makeatletter
\def\@tvsp{\mathchoice{{}\mkern-4.5mu}{{}\mkern-4.5mu}{{}\mkern-2.5mu}{}}
\def\llangle{\langle\@tvsp\langle}
\def\bigllangle{\big\langle\@tvsp\big\langle}
\def\rrangle{\rangle\@tvsp\rangle}
\def\bigrrangle{\big\rangle\@tvsp\big\rangle}
\def\Bigllangle{\Big\langle\@tvsp\Big\langle}
\def\Bigrrangle{\Big\rangle\@tvsp\Big\rangle}
\def\lllangle{\langle\@tvsp\langle\@tvsp\langle}
\def\biglllangle{\big\langle\@tvsp\big\langle\@tvsp\big\langle}
\def\rrrangle{\rangle\@tvsp\rangle\@tvsp\rangle}
\def\bigrrrangle{\big\rangle\@tvsp\big\rangle\@tvsp\big\rangle}
\def\Biglllangle{\Big\langle\@tvsp\Big\langle\@tvsp\Big\langle}
\def\Bigrrrangle{\Big\rangle\@tvsp\Big\rangle\@tvsp\Big\rangle}
\makeatother

\begin{document}
%%%%%%%%%%%%%%%%%%%%%%%%Basic  Stuff%%%%%%%%%%%%%%%%%%%%%%%%%%
{\author{\textsc{Bharath Ron}}
\title{{{\LARGE \bfseries \textsc{A Reconstruction of}}\\\vspace{0em}\huge\bfseries\textsc{{Algebraic Quantum Theory}}}\\\vspace{0em}}}
\date{}
\maketitle
\vspace{0em}

\begin{adjustwidth}{2em}{2em}
\begin{abstract}
\noindent We obtain a condensed reconstruction of algebraic quantum theory, emphasizing its foundational aspects and algebraic structure. We obtain the $W^*$-algebra structure from elementary assumptions about observers and how they can observe reality. This paper highlights the need for $W^*$-algebra structure by directly obtaining the mathematical axioms from simple thought experiments.\\\par

\noindent {\bfseries\textsc{Keywords:}} algebraic quantum theory, reconstruction, $W^*$-Algebras
\end{abstract}
\end{adjustwidth}

\vspace{2.5em}
\noindent Every scientific theory comprises of a \emph{linguistic representation} of a collection of postulates speculating the relations between facts of the world. The linguistic structures enable predictions about facts of the world by leveraging the postulated relations between the facts. The predictive capabilities of a scientific theory can be made more reliable by refining its faulty postulates. In this case we can only modify or remove the faulty postulates. Such refinements lead to theories with fewer postulates which are easier to validate and provide more reliable predictions. %We can hence quantify scientific progress in terms of the size of the smallest set of postulates that completely captures the scientific theory's predictive capabilities. 
Some of these faulty postulates might be embedded within the linguistic structures of the language used for postulate representation. To circumvent such linguistic issues, we may retain only those structures common to all languages. Such universally common linguistic structures are \emph{the rules of logic} and hence we can view \emph{mathematics as the universal language of postulate representation}. By using mathematics for postulate representation physics effectively avoids issues arising from suboptimal linguistic structures. The reliability of predictions of physical theories depends solely on its postulates. Formulation of physical theories usually involves finding an appropriate mathematical theory where its postulates \emph{can} be represented. If some of the structures of the mathematical theory are unnecessary for postulate representation, then such structures can constrain predictions that end up conflicting with observed facts. For example, commutativity in classical theories leads to conflicts with observed facts. Hence, it is crucial to figure out which assumptions of the mathematical theory are \emph{necessary} for representation of postulates. This necessity condition establishes an equivalence between the set of assumptions of the mathematical theory and the set of postulates of the corresponding physical theory. We say such a mathematical theory the \emph{universal theory} for the postulates.

Our objective is to develop the universal mathematical theory for the postulates we only consider the structures essential for postulate representation. The study of such peculiar structures may contribute to the development of new mathematics, which might be useful in the future. It is however not our focus here. 
We show the necessity of $W^*$-algebra structure for the postulates about how we observe the world around us, and hence may be viewed as a reconstruction of quantum theory.

\subsection*{Generalised Probabilistic Theories}\label{sec:gpt}
The physical systems of interest in quantum theories are those whose observed phenomena can be described by the notion of an \emph{observable}. The physical idea here is that the world can be described by sentences of the form \emph{the observable $\Sigma$ has the value $O$}. The next step is to articulate this idea in mathematical language. This means that we need to construct an appropriately predictive mathematical theory where sentences like the one above can be represented. Hence the starting point is the construction of a universal representation for this postulate. The construction of such a universal representation for the postulate we use here is closest to the instrumentalist approach, see \cite{Kraus},\cite{Ludwig1},\cite{Ludwig2}, with some ideas borrowed from the quantum logic approach, see \cite{Jauch},\cite{Kalmbach},\cite{Peres},\cite{Varadarajan},\cite{vonNeumann}, and algebraic quantum theory. This can be thought of as a modern formulation of Heisenberg's original idea. The starting point is the notion of measuring instruments and preparation instruments and since the only way to verify facts about the world is by performing measurements in the world, the approach is very general. Any changes occurring in the instruments during \emph{measurements} will be accepted as objective outcomes. From this point of view, the fundamental notions of quantum theory will have to be defined operationally in terms of measuring instruments, preparation procedures, and the prescriptions for their application. The theory is then interpreted entirely in terms of such instruments and outcomes which are the changes occurring to instruments.

Every preparation procedure is characterised by the kind of system it prepares. The measuring instruments are capable of undergoing upon performing a measurement evaluating the truth value of collection of possible outcomes. The possible results of such an experiment are called outcomes of the experiment and the observable change in the instrument is called an effect. To simplify the procedure consider instruments that record \emph{hits}. These instruments perform simple \emph{yes-no} measurements. Any measurement can be interpreted as a combination of \emph{yes-no} measurements. These \emph{yes-no} instruments can be used to build any general instrument. Suppose we have such a measuring instrument, label its measuring instrument by $O$. If the experiment is conducted a lot of times, we get a relative frequency of occurrence of \emph{yes}. Here \emph{yes} is an observable change in the instrument. It is hence an observable effect. To every preparation procedure $\rho$ and measuring instrument $O$ there exists a probability $\mu(\rho| O)$ of the occurrence of \emph{yes} associated with the pair,
$$(\rho,O)\longrightarrow \mu(\rho|O).$$
The numbers $\mu(\rho|O)$ are called operational statistics. 

Two completely different preparation procedures may give the same operational statistics, that is, they may give the same probabilities for every experiments. Such preparation procedures must be considered equivalent operationally. An equivalence class of preparation procedures yielding the same operational statistics for experiments is called an ensemble. Similarly, there may be two measuring instruments that have the same chances of undergoing a change for similarly prepared systems. Such measuring instruments must be operationally considered to be equivalent. An equivalence class of change for measuring instruments is called an effect. An effect is the equivalence class of all instruments that undergo a change for the same possible outcome. By considering equivalence classes we have obtained the structure of sets. Since we are considering equivalence classes, the construction and behaviour of the instruments are irrelevant to the discussion. We denote the set of ensembles by $\mathfrak{S}$ and the set of effects by $\mathfrak{E}$. The pairing describing the operational statistics is
$$\mathfrak{S}\times \mathfrak{E}\xlongrightarrow{\mu}[0,1].$$
For a possible outcome $O$, we will denote the corresponding effect by $\mu(\cdot|{O})$. Each effect acts on the ensembles of the system, and each ensemble acts on the effects of the system to yield the corresponding operational statistics
\begin{align*}
	\mu\big(\cdot|\:{O}\big)&:\:\rho \mapsto \mu(\rho|O),\\
	\mu\big(\:\rho\:|\cdot\big)&:\:O \mapsto \mu(\rho|O).
\end{align*}
The sets $\mathfrak{E}$ and $\mathfrak{S}$ together with the pairing $\mu(\cdot|\cdot)$ will be called an operational theory. Since any two preparations giving the same result on every effect represent the same ensemble and two measurement procedures that cannot distinguish ensembles represent the same effect. Therefore ensembles and effects must mutually separate each other respect to $\mu(\cdot|\cdot)$.

If we view the ensemble as a representation of the experimenter's beliefs about the universe, the collection of effects allow the experimenter to verify his beliefs about the universe. The results of experiments represent an updating of the experimenter's beliefs. If we define an observer as an equivalence class of experimenters with the same set of beliefs, then we may view operational theories \emph{as the theory of observers and the observed.} The set of ensembles corresponds to the set of all beliefs the observer \emph{can} hold, and represents the observer's \emph{subjective reality}. Since the space of effects is the only link between an experimenter and objective reality, we may identify the space of effects with objective reality itself. From this point of view, we may view operational theories as the theory of duality between the subjective and the objective.%The pursuit of science is functor categorical with respect to \emph{reality} itself.}
%%
%
%From here on, we will avoid unnecessary philosophy and verbal musings, and stay closest to reality by avoiding worldly languages and sticking to the universal language of mathematics.
\subsection*{Locally Convex Structure}\label{LocallyConvex}
In order to be able to work with effects and ensembles we now embed them inside sets with mathematical structure that respects the expected operational relations. We now describe how the collection of effects and the collection of ensembles inherit mathematical structures. In our case the operational requirement is that we should be able to make sense of taking mixtures of effects and ensembles. Suppose we can prepare systems in different ensembles by varying settings of a preparation procedures, then by adjusting the settings we should be able to prepare systems in mixtures of ensembles. We expect the space of ensembles to have the structures to describe mixtures. 

Let us assume that we can form such mixtures by changing the settings on the preparation procedures. If such adjustments are parametrised by real numbers, such as rotating a knob, then the notion of convexity needed for formalising comes from the space of real numbers. The space of real numbers is a convenient space because its convexity structure is related to a lot of other structures, such as its topology, its vector space structure, and so on. As discussed in \cite{Gudder}, such a notion of mixing corresponds mathematically to the notion of convex combination. Hence real vector space structure acts as a good starting point. Hence we may expect the set $\mathfrak{S}$ to have the structures for making sense of convexity and we expect that each functional $\mu(\cdot|O)$ preserves the convex structure since we expect the operational statistics to preserve this convexity. To make sense of taking convex linear combinations we need to be able to make sense of linear combinations. Hence we must embed the set of effects and the set of ensembles in a vector space with respect to a field that is at least as large as the real numbers. It is important to note that preparation procedures and measuring instruments producing the same ensembles and effects are not equal, in fact, the notion of equality will not even make sense. The transition from preparation procedures and measuring instruments to ensembles and effects is a transition from the real physical world to the abstract mathematical world. It should also be noted that it does not make sense to \emph{prepare} closed systems, one has to assume such systems start off in some state a priori.

A generalised probabilistic theory is an embedding of $\mathfrak{S}$ and $\mathfrak{E}$ inside vector spaces, such that the ensembles and effects are uniquely determined by the operational statistics they produce. This uniqueness of operational statistics is called the principle of tomography in quantum foundations and quantum information literature. To construct a generalised probabilistic theory for $\mathfrak{S}$ and $\mathfrak{E}$ we start by viewing effects and ensembles as linear functionals on a suitable space. Denote by $\mathcal{A}_*$ the set of maps
$$\mu(O)=\sum_{i\in F} \lambda_i \mu\big(\rho_i |O\big),$$
for every $O$ in $\mathfrak{E}$, where $\{\rho_i\}_{i\in F}$ is a finite collection of ensembles and $\lambda_i$ are scalars. Since the notion of convexity is inherited from the notion of convexity on the real line, we assume that the field $\mathbb{K}$ contains the field of real numbers. The function $|\cdot|:\KK\to \RR^+$ denotes the standard modulus function on $\KK$. Any linear combination of such maps also belongs to $\Acal_* $, it follows that $\Acal_* $ is a $\KK$-vector space. Denote by $\mathcal{E}$ the set of maps,
$$E(\rho)=\sum_{i\in F}\lambda_i \mu\big(\rho | O_i\big),$$
for every $\rho$ in $\mathfrak{S}$, where $\{O_i\}_{i\in F}$ is a finite collection of effects and $\lambda_i$ are scalars. It follows that $\mathcal{E}$ is also a $\KK$-vector space. We embed ensembles inside $\mathcal{A}_*$ with the map
\begin{align*}
	\rho \mapsto \mu\big(\rho\:|\cdot\big).
\end{align*}
Similarly, we embed effects in $\mathcal{E}$ with the map
\begin{align*}
	O \mapsto \mu\big(\cdot|\:O\big).
\end{align*}
The elements of the convex subsets $\mathfrak{E}$ of $\Ecal$ and $\mathfrak{S}$ of $\Acal_* $ are called effects and states respectively.

The vector space structure of $\Acal_* $ and $\Ecal$ allows us to develop algebraic tools for studying effects and ensembles. So we embedded an operational theory by viewing the effects and ensembles as linear functionals on each other and took \emph{quotients} with operational equivalences. Since we expect the ensembles and effects to be determined by the probabilities they produce we must also expect the vector spaces $\Ecal$ and $\Acal_* $ to inherit a relation between each other from the pairing $\mu(\cdot|\cdot)$ of $\mathfrak{E}$ and $\mathfrak{S}$. So we must have a pairing 
$$\big\langle\cdot|\cdot\big\rangle_\Ecal : \mathcal{A}_*\times \mathcal{E}\to \mathbb{K},$$
Such that, for every preparation procedure $\rho$ and measurement outcome $O$,
$$\Big\langle \mu(\rho|\cdot)\:\big|\:\mu(\cdot|O)\Big\rangle_\Ecal=\mu(O|\rho)$$
Hence we obtain a pairing of locally convex spaces equipped with the weak topologies. We now have access to the theory of conjugate duality of locally convex spaces.

%The deeper reason for the introduction of the complex number field is that it is algebraically closed and allows for factorisation. We will later see that the complex structure allows us mathematically to make sense of the physical idea that an instrument which undergoes a change for a collection of events can be replaced by a collection of instruments for those simultaneously measurable outcomes. Such a decomposition of a given effect into a collection of smaller, simultaneously measurable effects is made precise using complex structures by the spectral theory.

\subsection*{Banach Space Structure}
The operational requirements force the space of effects and ensembles to be vector spaces. The vector space structure is however insufficient for doing analysis with effects and ensembles which would allow the theory to be predictive. We now discuss how to introduce additional mathematical structures on the space of effects and ensembles, based on operational heuristics. We will later discuss the mathematically precise formulations of the heuristics.

Suppose we have two measuring instruments for an outcome where one is more accurate than the other. In such a case, the less accurate measuring instrument will readily undergo a change compared to the more accurate instrument. If the instrument is too accurate, it becomes difficult to observe any changes occurring in the measuring instrument. In this sense, the more accurate instrument should be \emph{closer} to the instrument which never undergoes any changes. So, we must expect $\Ecal$ to have some structure that quantifies how readily an instrument undergoes a change. The instrument which always undergoes a change should be the easiest to notice changes and the instrument which never undergoes any changes should be the hardest. Suppose we have two measuring instruments which undergo changes for outcomes $O$ and $O'$, and let $\mu(\cdot|O)$ and $\mu(\cdot|O')$ be the corresponding effects. For combined instruments, we expect the accuracy change for the combined instruments to depend on the accuracy of the component instruments. Hence the changes in accuracy must be expected to vary proportionally to the changes in accuracy of its components. This gives rise to continuity requirements on the space of effects. We assume the existence of the unit $1$, which corresponds to the instrument which is always true for any preparation procedure. The unit must be unique since any other such element must be operationally equivalent. {Operationally, the unit corresponds to the \emph{existence} of the system itself, since it must always be true.} Similarly we assume the existence of the unique $0$ element which outputs \emph{no} for every preparation procedure, and can be thought of as a faulty instrument which never undergoes any changes. Note that the less accurate instrument undergoes a change for more events than intended. A faulty instrument never undergoes any change, for any event. It has no possibility to be inaccurate, and hence correspond to the most accurate instruments.

For any collection of instruments for a given outcome we can obtain a relation between such instruments by their accuracy. Consider two instruments which can undergo changes under some \emph{similar} situations. This means that we may have some redundancies, or common inaccuracies, that is, they can measure some common possible outcomes, or alternatively, they have differing accuracy for measuring the common outcome. If the accuracy of either of the instruments is changed, the accuracy of the combined instrument should also change, and this change should be proportional to the change in accuracy of the instrument.

The notion of accuracy of instruments must give rise to a function on the space of effects. In our case, it is easier to model the inaccuracy. If changes in accuracy are modelled using the addition operation on the real line, then the notion of inaccuracy must correspond to a real-valued function $f$ such that,
$$f\big(\mu(\:\cdot\:|\:O\:)+ \mu(\:\cdot\:|\:O'\:)\big)\leqq f\big(\mu(\:\cdot\:|\:O\:)\big) +f\big(\mu(\:\cdot\:|\:O'\:)\big).$$
Such functions are called sub-linear maps. The notion of inaccuracy of instruments naturally gives rise to triangle inequality like properties.

We can describe this \emph{inaccuracy function} more conveniently as a supremum. Since inaccuracy represents the maximal likelihood of undergoing a change, it must correspond to the supremum of undergoing a change over all preparation procedures. That is to say, the accuracy in the real world can only depend on \emph{reality} which we expect to be closely related to preparation procedures. This gives us a definition in terms of supremum over preparation procedures, and can be defined on all of $\Ecal$,
$$\big\| E\big\|_{\Ecal}\coloneqq \sup_{\omega\in\mathfrak{S}}\Big[\big| \big\langle E|\omega\big\rangle_\Ecal\big|\Big],$$
for every $E$ in $\Ecal$, and the supremum is taken over all preparation procedures and which corresponds to the convex embedded subset $\mathfrak{S}$ of $\Acal_* $. It immediately follows that $\|\cdot\|_\Ecal$ satisfies the triangle inequality, and scaling condition,
\begin{align*}
	\|E +F\|_\Ecal &\leqq \|E\|_\Ecal+\|F\|_\Ecal.\\
	\big\|\lambda E\big\|_\Ecal&=|\lambda| \big\|E\big\|_\Ecal,
\end{align*}
for all $E,F\in \Ecal$ and $\lambda$ is a scalar. By definition of the norm we have $\|1\|_\Ecal=1$. By the assumption that there will always exist some preparation procedure with non-trivial operational statistics, we deduce that $\|\cdot\|_\Ecal$ is non-trivial, that is, $\|E\|_\Ecal=0$ if and only if $E\equiv 0$. Hence $\|\cdot\|_\Ecal:\Ecal\to \RR$ defines a norm on $\Ecal$.

Similarly, $\Acal_* $ also inherits operator norm from $\Ecal$. We can have two preparation procedures, where the preparations settings are slightly varied. We can say two ensembles are close to each other if the operational statistics are close to each other for all effects. Hence how similar two preparation procedures are can also be described as a supremum, and this can be described on all of $\Acal_* $. 
$$\big\| \omega \big\|_{\Acal_* }\coloneqq \sup_{E\in \mathfrak{E}}\Big[\big| \langle E|\omega\rangle_\Ecal\big|\Big],$$
for every $\omega$ in $\Acal_*$ and the supremum is taken over all possible measurements which corresponds to the convex embedded subset $\mathfrak{E}$ of $\Ecal$. We can hence require both $\Ecal$ and $\Acal_* $ to be normed vector spaces.

The notion of limits is a mathematical notion, and by assumption must be common to every experimenter, since we expect the rules of logic to be a universal property of languages. This ensures that the predictions obtained by an experimenter assuming the existence of limits must remain valid for any other experimenter. This allows the experimenter to mathematically analyse and predict facts about the world based on the experimentally verified beliefs. Hence we may assume that the normed spaces $\Ecal$ and $\Acal_*$ are complete with respect to their respective norms, and $\mathfrak{E}$ and $\mathfrak{S}$ are closed convex subsets of $\Ecal$ and $\Acal_*$ respectively. The spaces $\Ecal$ and $\Acal_* $ are generated by limits of linear combinations of elements of $\mathfrak{E}$ and $\mathfrak{S}$ respectively. By the boundedness of operational statistics we note that effect and ensembles give rise to norm bounded linear functionals on $\Acal_* $ and $\Ecal$ respectively. By assumption $\mu(\cdot|\cdot)$ separates $\mathfrak{S}$ and $\mathfrak{E}$. Since $\langle \cdot|\cdot\rangle_\Ecal$ coincides with $\mu(\cdot|\cdot)$ for ensembles and effects, it follows by the density of the span of $\mathfrak{E}$ that
$$\big\langle \cdot|\cdot\big\rangle_\Ecal: \Acal_*\times \Ecal \to \mathbb{K},$$
is a non-degenerate pairing, and hence $\langle \Ecal|\Acal_* \rangle_\Ecal$ must be a dual pair. If $(\Acal_* )^*$ is the space of norm bounded linear functionals on $\Acal_* $ and as Banach spaces it follows that
$$\Ecal\cong \big(\Acal_* \big)^*$$
This gives us access to the tools of duality theory of Banach spaces for making predictions.
\subsection*{Banach Algebra Structure}
We expect varying accuracy of either of the measuring instruments corresponding to $E$ or $F$ to vary the accuracy of $EF$ accordingly. Hence we expect the algebraic structure on $\Ecal$ to respect the topological structure on $\Ecal$ coming from the norm $\|\cdot\|_\Ecal$. We may perform a measurement before or after performing some other measurement. This composite measurements can itself be thought of as a measuring instrument. Heuristically, when these concatenated measurements have inaccuracies which are independent of each other, we must expect the accuracy of the concatenation, to amplify the inaccuracies so the total inaccuracy would be that of the first instrument \emph{and} the second instrument and hence must correspond to the product of inaccuracies. This can only be the case when the accuracies of the two instruments are independent. Hence the accuracy of the concatenation must be expected to be less than the product of individual accuracies. This gives rise to an algebra structure on the space of effects (note that there is no postulate so far \emph{requiring} commutativity of the product structure). The space $\Ecal$ comes equipped with a natural left and right actions of effects on it given by
$$\Lcal_E F=EF\:\:;\:\:\Rcal_F E=EF.$$
We now study the topological properties of the product.
 
The composite element $\Lcal_EF=EF$ corresponds to a well-defined measuring instrument, and the accuracy of the combined instrument is given by $\|EF\|_\Ecal=\sup_{\omega\in \mathfrak{S}}|\langle EF|\omega\rangle_\Ecal|$. Since we are allowed to apply $E$ after applying any other measuring instrument the accuracy of left-composition by $E$ is given by,
\begin{align*}
	\big\|\Lcal_E\big\|_{\Bcal(\Ecal)}=\sup_{F\in\mathfrak{E}\atop\omega\in\mathfrak{S}} \Big[\big|\langle EF| \omega\rangle_\Ecal\big|\Big]=\sup_{F\in \mathfrak{E}}\bigg[\sup_{\omega\in \mathfrak{S}}\Big[\big| \langle EF|\omega \rangle_\Ecal\big|\Big]\bigg]=\sup_{F\in \mathfrak{E}}\Big[\big\| EF\big\|_\Acal\Big].
\end{align*}
Similarly, the norm of $\Rcal_F$ is given by,
$$\big\|\Rcal_F\big\|_{\Bcal(\Ecal)}=\sup_{E\in \mathfrak{E}}\Big[\big\| EF\big\|_\Ecal\Big].$$
So, both $\Lcal_E$ and $\Rcal_F$ belong to the space $\Bcal(\Ecal)$ of bounded operators on $\Ecal$.

Since we expect the accuracy of the composite to change proportionally to the change in accuracy of the composing instrument, we expect both $\Lcal_E$ and $\Rcal_F$ to be continuous. This must relate the norm $\|\cdot\|_{\Bcal(\Ecal)}$ and the norm $\|\cdot\|_\Ecal$. To relate the two norms we start with the following relation; $\Rcal_FE=EF=\Lcal_E F.$ By continuity of $\Rcal_F$ we have, $\| \Rcal_F E\|_\Ecal =\| \Lcal_E F\|_\Ecal\leqq \| \Lcal_E \|_{\Bcal(\Ecal)}\:\| F\|_\Ecal.$ Hence $\{\Rcal_F E\}_{F\in \mathfrak{E}}$ is a bounded set for every $E$. 

Assuming $\Ecal$ is a Banach space, by the Banach-Steinhaus theorem or the uniform boundedness principle, point-wise boundedness of the set $\{\Rcal_F E\}_{F\in \mathfrak{E}}$ implies that $\{\Rcal_F\}_{F\in {\mathfrak{E}}}$ is uniformly bounded, with the uniform bound $\vertiii\Rcal$. Hence for every $E\in \Ecal$,
$$\big\|\Rcal_F\big\|_{\Bcal(\Ecal)} \leqq \vertiii\Rcal \:\big\|F\big\|_\Ecal,$$
The map $\Rcal$ is a bounded linear map,
\begin{align*}
	\Rcal:\:&\Ecal\to \Bcal(\Ecal)\\
	&F\to \Rcal_F.
\end{align*}
The mapping $\Rcal$ continuously maps the Banach space ${\Ecal}$ into the Banach algebra $\Bcal({\Ecal})$ of bounded linear maps on $\Ecal$. The mapping $\Rcal$ is a continuous algebra homomorphism. Assuming $\|1\|_\Ecal=1$, we have,
$$\big\|E\big\|_\Ecal=\big\|\Rcal_E(1)\big\|_\Ecal \leqq \big\|\Rcal_E\big\|_{\Bcal(\Ecal)}\leqq \vertiii\Rcal \:\big\|E\big\|_\Ecal.$$
We also have $\|\Rcal_1 (1)\|=\|1\|=1$, hence it follows that $\vertiii \Rcal =1$.

Hence $\Rcal(\Ecal)$ is a norm-closed sub-algebra of $\Bcal({\Ecal})$. The map $\Rcal$ is an algebraic isomorphism from ${\Ecal}$ to $\Rcal({\Ecal})$. Since for any two bounded linear operators $S$ and $T$ on ${\Ecal}$ we have, $\|TS\| \leqq \|T\|\:\|S\|$, $\Rcal({\Ecal})$ inherits this property and we have,
$$\big\|EF\big\|_\Ecal\leqq \big\|E\big\|_\Ecal\big\|F\big\|_\Ecal$$
Since $\|\Rcal_1\|_{\Bcal(\Ecal)}=1$ we have, $\|\Rcal_E(1)\|_\Ecal =\|E\|_\Ecal \leqq \|\Rcal_E\|_{\Bcal(\Ecal)} \leqq \|\Rcal_1\|_{\Bcal(\Ecal)}\:\|E\|_\Ecal=\|E\|_\Ecal.$ Similarly for $\Lcal$.  Then we must have
$$\big\|\Rcal_E\big\|_{\Bcal(\Ecal)}=\big\|E\big\|_\Ecal=\big\|\Lcal_E\big\|_{\Ecal}.$$
Hence the embedding,
$$\Rcal:\Ecal\rightarrowtail \Rcal(\Ecal)$$
is an isometric isomorphism to its image and $\Rcal({\Ecal})$ is a Banach algebra. The space of states acts as continuous linear functionals on $\Rcal(\Ecal)$, given by the map $\Rcal_E\mapsto \big\langle \omega|E\big\rangle_\Ecal,$ with $$|\omega(\Rcal_E)|=|\langle \omega|E\rangle_\Ecal|\leqq \|\Rcal_E\|_{\Bcal(\Ecal)}$$
hence the elements of $\Acal_* $ respect the topological properties of $\Rcal(\Ecal)$. %This gives us access to the tools of the algebraic duality between ideals and subspaces.

We may now note what it means to re-perform an experiment or attempt to measure the same outcome. For an outcome $R$, the preparation procedure concatenated by a measuring instrument corresponding to the effect $\mu(\cdot|R)$ is itself a preparation procedure. To such preparation procedure, an application of the same measuring instrument again should not change the ensemble the procedure produces, since the the belief about outcome has already been verified. Since the concatenation of measurements corresponds to the product structure of $\Rcal(\Ecal)$, it must be the case that whenever $E$ in $\Rcal(\Ecal)$ corresponds to an experimental outcome it must correspond to a projection element
$$E^2=E.$$
The effects of measuring outcomes correspond to orthogonal projections in $\Rcal(\Ecal)$.
\subsection*{Complex Structure}
As soon as $\Rcal(\Ecal)$ possesses an algebra structure, we can make sense of polynomials in $\Rcal(\Ecal)$. 
Suppose a measuring instrument consists of a collection of mutually compatible instruments, that is, the measurements performed by such instruments do not affect the outcomes of the other instruments. Suppose the instrument undergoes a change if it undergoes a change for each of its component instruments or sub-instruments, then the probability of outcome for such an instrument should correspond to product of probabilities of its components. 

Let $E$ be the effect corresponding to such an instrument and $E_i$ be the effects corresponding to its components. Then for any ensemble $\omega$ we must have,
$$\big\langle \omega|E\big\rangle_\Ecal=\prod_{i\in \Ical}\:\langle \omega|E_i\rangle_\Ecal.$$
Since each effect corresponds to an equivalence class of measuring instruments, we may expect there to exist some measuring instrument which is the collection of compatible measuring instruments which cannot be further divided. This allows us to postulate that \emph{it is always possible to decompose a measuring instrument into its maximally divided collection of sub-instruments}. We assume that it is always possible to decompose a measuring instrument as in terms of its component sub-instruments. To make sense of this we need structures that enable us to factorise polynomials in $\Rcal(\Ecal)$. By the fundamental theorem of algebra, we can always factorise polynomials in the field of complex numbers. Hence we may assume the underlying field to be that of complex numbers
$$\KK\equiv \CC.$$
Therefore we assume the pairing to be a real bilinear map to the field of complex numbers
$$\big\langle \cdot|\cdot\big\rangle_{\Rcal(\Ecal)} :\Acal_* \times \Rcal(\Ecal)\to \CC.$$
For mathematical convenience, we will transfer this complex structure to the vector spaces $\Acal_* $ and $\Rcal(\Ecal)$ themselves, and assume that $\Acal_* $ and $\Rcal(\Ecal)$ are complex vector spaces. We require the scaling of both effects and ensembles by a unit length complex number to give the same pairing as the unscaled pairing. This requirement along with real linearity in each argument forces us to assume the pairing is sesquilinear. We now have access to the theory of complex numbers, and complex analysis.

Since the set of quaternions is not an algebra over $\CC$, they do not intrinsically allow for factorisation as required in the discussion above. On the other hand, the structures in quaternions are not \emph{required} by the postulates so far, and hence quaternions cannot be part of the universal mathematical theory for these postulates. We may therefore rule out quantum mechanics in the field of quaternions. 

\subsection*{$W^*$-Algebra Structure}
We now consider operations on the space of effects that are physically meaningful. Suppose we perform a rearrangement or readjustments of all the measuring instruments. Then we can consider all the measuring instruments which have been turned to non-functioning instruments, these instruments never undergo any changes, and the re-arrangement or re-adjustment of instruments acts as a switch-off for those instruments. Assuming such manipulations keep the system within itself, it must keep the space of effects invariant. However the only way to find out what changes we have made to the system is by performing experiments on the system. Since the space of states is separating for the space of effects, any manipulation we do to the space of effects are \emph{seen} by the space of states. Hence we consider the equivalence class of such manipulations on the space of effects, which affect the pairing $\langle \cdot| \cdot\rangle_{\Rcal(\Ecal)}$ of $\Rcal(\Ecal)$ with $\Acal_* $ in the same way. Each such equivalence class of manipulation of the system corresponds to a linear map,
$$T:\Rcal(\Ecal)\to \Rcal(\Ecal).$$
The manipulation $T$ of the system can also be thought of as a manipulation of the preparation procedures. We can also consider the equivalence class of manipulations of the system giving rise to the same change in preparation procedures, and for every ensemble $\omega$ this can be described in terms of operational statistics by the map
$$R_T \omega:E\mapsto\big\langle \omega|TE\big\rangle.$$
This map of ensembles can be extended to $\Acal_* $ by linearity, such that for every preparation procedure $\rho$ and $\sigma$ and scalars $\alpha$ and $\beta$,
$\alpha \mu(\rho|\cdot)+ \beta \mu(\sigma|\cdot)\mapsto \alpha\mu\big( T(\cdot)| \rho\big)+\beta\mu\big(T(\cdot)|\sigma\big).$
Hence to each physically meaningful manipulation of the system there must exist a bounded linear map,
$$R_T:\Acal_* \to \Acal_* ,$$
This map on $\Acal_*$ gives us a map,
$$T^\dagger:\Rcal(\Ecal)\to \Rcal(\Ecal).$$
It is determined by the pairing $\langle \cdot|\cdot\rangle_{\Rcal(\Ecal)}$ of $\Rcal(\Ecal)$ and $\Acal_* $ by,
$$\big\langle T^\dagger \omega|E\big\rangle_{\Rcal(\Ecal)}=\big\langle R_T\omega|E\big\rangle_{\Rcal(\Ecal)}$$
Note that $R_T$ can also be thought of in terms of the left-action $\Lcal$, since it corresponds to the transformation by $T$ of effects which can be pre-composed with effects. Hence the $\dagger$-operation would have the effect of reversing the order of taking product, when viewed in terms of ensembles. This can be described more directly in terms of the pairing $\langle\cdot|\cdot\rangle_{\Rcal(\Ecal)}$.

By the sesquilinearity of the pairing $\langle \cdot|\cdot\rangle_{\Rcal(\Ecal)}$ it follows for all physically relevant operators $T,S$ on the space $\Rcal(\Ecal)$ and for all scalars $\lambda$, and $\mu$ that,
\begin{align*}
	({\lambda T+\mu S})^\dagger &=\overline{\lambda} T^\dagger +\overline{\mu} S^\dagger , \\
	(TS)^\dagger&=S^\dagger T^\dagger.
\end{align*}
The operation $\dagger:\Rcal(\Ecal)\to \Rcal(\Ecal)$ seems to be completely algebraic in nature, however it also  possesses interesting topological properties, since the topologies on the space of effects and state space operationally correspond to the physical notion of accuracy. In order to describe the topological properties of the operation $\dagger$, we now describe the notion of complementary effect and its relation to the notion of topological complementation on Banach spaces. 

Suppose we have a measuring instrument which can measure an outcome $R$, the change occuring in the instrument corresponds to the effect for $R$. If the measuring instrument does not undergo any change, then we can infer that the outcome $R$ had not occurred. Since the outcome $R$ not occuring is also an outcome by itself, denoted by $\neg R$, and the same measuring instrument also be used for measuring $\neg R$, the instrument not undergoing a change corresponds to the effect for $\neg R$. Hence we must have an operation on the space of effects which assigns to every effect $\mu(\cdot|R)$ corresponding to the outcome $R$, the effect $\mu(\cdot|{\neg R})$ corresponding to the outcome $\neg R$. We call $\mu(\cdot|{\neg R})$ the complement of $\mu(\cdot|R)$. Generally, we denote the complement of $E$ by $E^\perp$. We obtain a mapping on the space of effects. This notion of complement of effects is closely related to the notion of topological complementation of closed subspaces on Banach spaces.

The Banach space structure on $\Rcal(\Ecal)$ allows us to make sense of closed subspaces of $\Rcal(\Ecal)$, with respect to the norm $\|\cdot\|_{\Bcal(\Ecal)}$ inherited from $\Bcal(\Ecal)$. Let $\Mcal$ be a closed subspace of $\Rcal(\Ecal)$, we say that $\Mcal$ is topologically complemented if there exists a closed subspace $\Mcal^\perp$ of $\Rcal(\Ecal)$ such that
$$\Mcal\oplus \Mcal^\perp \equiv \Rcal(\Ecal).$$
In such a case every vector $A$ in $\Rcal(\Ecal)$ can be uniquely decomposed as a direct sum of vectors $A_\Mcal+A_{\Mcal^\perp}$ where $A_\Mcal$ is in $\Mcal$ and $A_{\Mcal^\perp}$ is in $\Mcal^\perp$. Hence we can consider the projection map on $\Rcal(\Ecal)$,
$$E_\Mcal:\Rcal(\Ecal)\to \Rcal(\Ecal)$$
which sends each $A$ in $\Rcal(\Ecal)$ to the vector $A_\Mcal$ in $\Mcal$. Since we do not have to decompose $A_\Mcal$ any further we must have $E_\Mcal=E_\Mcal^2$.

The closed graph theorem suggests that the continuity properties of operators on Banach spaces and the topology of the product spaces are closely related, and says that an operator is continuous if and only if its graph is closed. Since both $\Rcal(\Ecal)$ and $\Mcal$ are closed, it follows that $E_\Mcal$ has a closed graph, and hence the projection,
$$E_\Mcal:\Rcal(\Ecal)\to \Rcal(\Ecal),$$
onto the closed complemented subspace $\Mcal$ of $\Rcal(\Ecal)$ is a continuous operator on $\Rcal(\Ecal)$. Conversely, every continuous projection $E_\Mcal$, corresponds to a unique closed complemented subspace $\Mcal$. Since the space of effects is contained in $\Rcal(\Ecal)$ and since each effect correspond to a continuous projection operator, it follows that every effect corresponds to a unique closed complemented subspace of $\Rcal(\Ecal)$, hence the notation for the complement effect $E^\perp$ is closely related to the notion of topological complementation on $\Rcal(\Ecal)$.

Every laboratory can be thought of as a subsystem of the universe, and can equivalently be described in terms of the preparation procedures it can prepare, or by all the experiments which we can perform in the laboratory. By switching off the laboratory, we switch off all the experiments we could perform in the laboratory, or equivalently switch of all the preparation procedures in the laboratory. The laboratory itself can be thought of as a measuring instrument which is compatible with all the experiments it can perform, and for the laboratory itself, this must correspond to the identity element. Hence to each laboratory, there must exist a projection element, which commutes with every experiment that can be performed in the laboratory. To each preparation procedure in the laboratory, this projection element corresponds to the switching on of the procedure. For each preparation procedure, we have a collection of laboratories in which the preparation procedure can exist. For example, a preparation procedure in a laboratory, is also a preparation procedure in a bigger laboratory which contains the smaller laboratory. The projection element corresponding to the smallest laboratory to which the preparation procedure can belong to, will be called the support of the preparation procedure. Hence we obtain a correspondence between projection elements and collection of preparation procedures for which the projection corresponds to the minimal laboratory. If $\mu(\rho|\cdot)$ is the ensemble corresponding to some preparation procedure $\rho$, then we denote by $E_\rho$ the projection element as described above. We must require that
$$\big\|\mu\big(\rho\:|\cdot\big)\big\|_{\Acal_* }=\big\| \mu\big(\cdot|E_\rho\big)\big\|_{\Ecal}.$$
By linearity, the operator $(T^\dagger)^\dagger$ is determined by the pairing  $\mu(\rho|(T^\dagger)^\dagger(E))=\mu(R_T\rho|E)=\mu(\rho|TE)$ for all effects $E$ belonging to the support of $\mu_\rho$, and hence also for $E_\rho^\perp$. By the uniqueness of the pairing of closed complemented subspaces with its complement, and since ensembles and effects separate each other, we must require that
$$(T^\dagger)^\dagger=T.$$
For all physically relevant transformations $T$, the mapping $\dagger$ acts as an involution. The uniqueness of the topological complementation ensures that the operation $\dagger$ is an involution. 

The space of effects $\Rcal(\Ecal)$ is a subset of this collection, and hence we can consider this space of physically relevant operators on the space of effects itself as the object of study. Abusing notation we denote this algebra of operators by $\Rcal(\Ecal)$. From the discussion so far, we expect $\Rcal(\Ecal)$ to be a Banach algebra equipped with an involution
$$\dagger:\Rcal(\Ecal)\to \Rcal(\Ecal).$$
Such Banach algebras are called involutive Banach algebras, or Banach$^*$-algebras.

However ensembles are themselves closely related to effects in an opposite way via the notion of support. Hence the composite manipulation of the system acts in an \emph{independent} way on the space of effects and the state space. From the discussion about norms on $\Rcal(\Ecal)$ and $\Acal_* $, we know that the norms are closely related to the notion of accuracy of measuring instruments and preparation procedures. Since the manipulations introduced on the space of effects and state space are opposite of each other they are independent, and we should expect the accuracies of effects and ensembles to be multiplicative. The relation between support properties of ensembles and effects captures the complementary nature of such a composite manipulation $T^\dagger T$ of the system.

We now make this precise, via topological complementations and the expected relationship between subspaces of the state space and ideals of the space of effects. The norm of $T^\dagger T$ is given by
$$\big\| T^\dagger T\big\|=\sup_{E\in \mathfrak{E}\atop\omega\in \mathfrak{S}}\Big[ \big|\big\langle R_T\omega|TE\big\rangle_{\Rcal(\Ecal)}\big| \Big].$$
We denote the space ${\Bcal(\Rcal(\Ecal))}$ of bounded operators on $\Rcal(\Ecal)$. We must have,
\begin{align*}
	\big\| T^\dagger T\big\|=\sup_{E\in \mathfrak{E}\atop\omega\in \mathfrak{S}}\Big[ \big|\big\langle R_T\omega|TE\big\rangle_{\Rcal(\Ecal)}\big| \Big]=\sup_{E\in \mathfrak{E}}\Big[\big\|TE\big\|\Big]\sup_{\omega\in \mathfrak{S}}\Big[\big\|R_T\omega\big\|\Big]=\big\| T\big\|^2.
\end{align*}
The reason for the equality of products of supremums is due to the fact that for every $\omega$ we can split the state space as direct sum of complemented subspaces as 
$\Acal_* =(E_{R_T\omega})\Acal_* \oplus ({E_{R_T\omega}}^\perp) \Acal_*$ and the pairing vanishes if $TE$ does not belong to the support of $R_T\omega$, we can assume that the supremum is attained only when $TE$ belongs to the ideal $({E_{R_T\omega}}^\perp) \Ecal$. Hence we have,
$$\big\|T^\dagger T\big\|=\big\|T\big\|^2.$$
Banach$^*$-algebras satisfying this identity are called $C^*$-algebras. The following composite map gives us an embedding of $\mathfrak{E}$ inside ${\Bcal(\Rcal(\Ecal))}$,
$$R\mapsto \mu(\cdot|R)\mapsto \Rcal_{\mu(\cdot|R)}\mapsto \Rcal_{\Rcal_{\mu(\cdot|R)}}.$$
Note that it is sufficient to consider only the bounded operators, and any other unbounded yet meaningful manipulations can be described as maps on the bounded operators. The algebra of observables $\Acal$ is the collection of all elements of $\Bcal(\Rcal(\Ecal))$ which are bounded and correspond to physically meaningful manipulations of the system. We must have a representation of measuring instruments as an embedding inside the algebra of observables. Since any changes from such manipulations must be noticed by the space of ensembles, it follows that $\Acal_*$ remains separating for $\Acal$. Hence we must have a dual pairing,
$$\big\langle\cdot|\cdot\big\rangle:\Acal_*\times \Acal\to \CC.$$
Such that,
$$(\Acal_*)^*\cong \Acal$$
A $C^*$-algebra $\Acal$ which is a dual Banach space of $\Acal_*$ is called a $W^*$-algebra, and in such a case the Banach space $\Acal_*$ is called the predual of $\Acal$.

We may expect any operational theory to be represented by the pairing of a $W^*$-algebra, and its predual. By Sakai's predual theorem, see \cite{Sakai}-\cite{TomiyamaII}, $W^*$-algebras are the same as von Neumann algebras on a Hilbert space. Hence every quantum theory naturally requires the structures of Hilbert spaces and bounded operators on Hilbert spaces.

\subsubsection*{Philosophical Digression: Subjectivity vs Objectivity}
If we view ensembles as a representation of an experimenter's belief about reality, the collection of effects allow the experimenter to verify his beliefs about the universe. The results of experiments represent an updating of the experimenter's beliefs. If we define an observer as an equivalence class of experimenters with the same set of beliefs, then we may view operational theories \emph{as the theory of observers and the observed.} The set of ensembles corresponds to the set of all beliefs the observer \emph{can} hold, and represents the observer's \emph{subjective reality}. Since the space of effects is the only link between an experimenter and objective reality, we may identify the space of effects with objective reality itself. From this point of view, we may view operational theories as theories of duality between the subjective and the objective.

On the other hand, each physical theory can itself be thought of as an observation about how observers observe reality. If we expect there to exist some physical theory that allows us to predict about the physical world to any level of accuracy then reality can be thought of as a limit of physical theories. Then \emph{reality is a functor categorical object}, corresponding to the limit of physical theories in the convenient complete category of functors from the category of physical theories to the category of sets. Where the category of physical theories consists of physical theories as objects, and relation between physical theories as morphisms. Physical theories themselves act as \emph{higher effects} and  physics itself acts as \emph{higher reality}, and mathematics may itself thought of as the collection of \emph{higher ensembles}.

\subsubsection*{Acknowledgements}
I extend my sincere thanks to my master's advisor, Dr. G\"unther H\"ormann, for their invaluable guidance and support during my research. Their expertise and encouragement have been crucial to the completion of this work.

{\phantomsection\addcontentsline{toc}{chapter}{{Bibliography}}}

\end{document}